\documentclass[aps,prd,onecolumn,groupedaddress,showpacs,nofootinbib,amssymb]{revtex4-2}
\usepackage[dvips]{graphicx}
\usepackage{amssymb}
\usepackage{amsmath}
\usepackage{float}
\usepackage{graphicx,,color}
\usepackage{amsfonts}
\usepackage{bm}
\usepackage{cancel}
\usepackage{comment}
\usepackage{floatflt}
\usepackage{slashed}

\newcommand\be{\begin{equation}}
\newcommand\ee{\end{equation}}

\allowdisplaybreaks[4]

\begin{document}
\sloppy

\title{Primordial Cosmology of an Emergent-like Universe from
Modified Gravity: Reconstruction and Phenomenology Optimization with a Genetic Algorithm}
\author{V.K. Oikonomou,$^{1,2}$}\email{voikonomou@gapps.auth.gr;v.k.oikonomou1979@gmail.com}
\author{Gregory K. Kafanelis,$^{1}$}\email{gkafanel@auth.gr}
\affiliation{$^{1)}$Department of Physics, Aristotle University of
Thessaloniki, Thessaloniki 54124, Greece \\ $^{2)}$L.N. Gumilyov
Eurasian National University - Astana, 010008, Kazakhstan}

 \tolerance=5000

\begin{abstract}
General relativity has been the dominant paradigm theory of
gravity for the last century, it manages to explain a plethora of
astronomical and cosmological observations with high accuracy. A
major challenge faced by general relativity is, that it predicts a
singularity as the beginning of our universe. In this letter we
shall explore a well known idea in the literature, that of the
emergent universe and it's viability according to Planck 2018 data
and BBN constraints.
\end{abstract}


\maketitle

\section{Introduction}

Starting off there must be a discussion as to what an emergent
universe model encompasses. In their paper George F R Ellis and
Roy Maartens \cite{Ellis:2002we}, describe an initial state of the
universe in the infinite past, that inflates for a long period of
time (in their analysis that inflationary phase lasted an infinite
time), and after exiting the inflationary and reheating stages,
emerges our own universe. That scenario has some very interesting
and desirable characteristics. First off, it does not present a
horizon problem as the initial state is an Einstein static
universe, and thus matter and radiation had enough time to be in
casual contact. Secondly, if the initial size is sufficient the
universe never enters a Planckian epoch and consequently there is
no need for quantum gravity, as the scale of interest is within
the bounds of General Relativity. Lastly, it provides an elegant
solution to the singularity problem mentioned in the abstract. The
universe never had to experience a bouncing cosmology, as it
inflated from its initial state, so no singularity occurred. \par

In their original work, the authors assumed a positive curvature
universe and a scale factor constructed of exponential functions,
to instigate their further research into emergent universe
scenarios. We shall try to reconstruct a viable gravity model that
produces a generalized version of their scale factor, all while
assuming the mainstream proposition of a flat universe. The
general form of the scale factor that will be used throughout this
paper is:
\begin{equation}\label{scalefactor}
    a(t)=b \left(\exp \left(-\frac{\sqrt{2} t}{b}\right)+1\right)^n ,
\end{equation}
with b being positive and n negative, dimensionless quantities. It
can be easily seen that for negative time, the scale factor of the
universe is practically constant and approximately equal to
$\varepsilon$, with epsilon being a small positive  quantity. The
scale factor is asymptotically equal to zero for negative time. To
have the appropriate units, the argument t/b should be
t/(b*$\sqrt{\kappa}$), but as the analysis would be only
phenomenological, the constant $\kappa$, shall be set to unity.A
lot of ways to realize an emergent universe scenario have been
explored in the literature, but in the scope of this paper we
shall proceed with the reconstruction of an $f(R)$ gravity that
gives rise to this particular scale factor. Albeit, the common
factor used in such scenarios is of the form,
$b(1+e^{t/b})^{n}$,\cite{Verlinde:2016toy}, these kinds of factor
are not viable for matter-free reconstruction due to the emergence
of imaginary expressions, which are unnatural. The inflationary
phase of the universe ends naturally, with a finite number of
e-folds, resulting in a subsequent reheating phase.\par

Before we delve deeper into the topic of emergent universe
scenarios,  it is important to address a common criticism leveled
against them,  which is related to the issue of fine-tuning.
Although there are no scientific arguments that prevent the
universe to have occurred as a result of precise tuning in its
initial parameters, there exists a disdain in the cosmology
circles regarding such hypotheses as unsatisfactory and
fallacious, as they try to mask our ignorance about some yet
discovered physical mechanism. Without getting ahead of ourselves,
we must state that the current model reproduces the Planck 2018
observational data, for a wide range of its free parameters. In
light of that fact, we want to draw the conclusion, that emergent
universe models do not need fine tuning to be viable, and feasible
models with explanatory power exist, it is just that they are a
lot harder to find.
\begin{figure}[h]
    \centering
    \includegraphics[width=9cm]{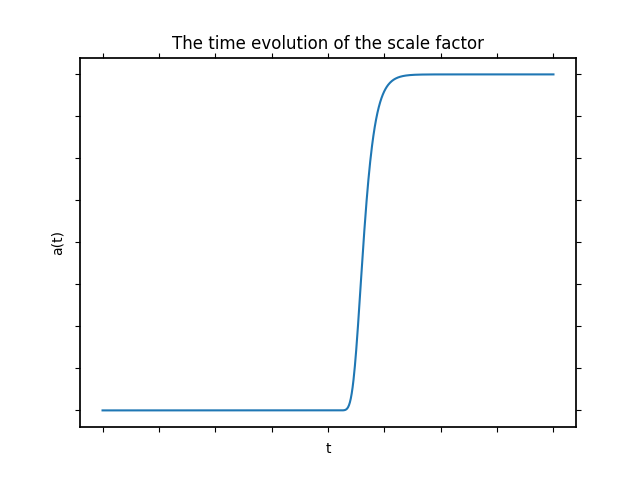}
    \caption{The scale factor in terms of time , for n=-60 and b=2000}
\end{figure}

Inflation \cite{inflation1,inflation2,inflation3,inflation4} is
now put into focus from most current and forthcoming experiments.
The aim is to pinpoint the B-modes directly experimentally, and
this is the central focus in the stage 4 Cosmic Microwave
Background (CMB) experiments
\cite{CMB-S4:2016ple,SimonsObservatory:2019qwx}. Apart from the
B-modes, there are indirect ways to detect an inflationary era,
for example the detection of a small (or negligible) anisotropy
stochastic gravitational wave background
\cite{Hild:2010id,Baker:2019nia,Smith:2019wny,Crowder:2005nr,Smith:2016jqs,Seto:2001qf,Kawamura:2020pcg,Bull:2018lat,LISACosmologyWorkingGroup:2022jok}.
These future gravitational wave experiments will directly probe
the inflationary modes that became subhorizon after the
inflationary era. Apart from the LIGO-Virgo successes and exciting
observations, recently the NANOGrav collaboration confirmed the
well anticipated stochastic gravitational wave detection
\cite{nanograv}, which was also confirmed by other pulsar timing
arrays (PTA) experiments
\cite{Antoniadis:2023ott,Reardon:2023gzh,Xu:2023wog}. The
scientific community was almost certain that the 2020 signal
detection was not due to the actual pulsar red noise but due to a
stochastic gravitational wave background, which was confirmed in
2023 due to the presence of Hellings-Downs correlations. After the
2023 NANOGrav announcement, a large stream of research articles
were produced, trying to explain the signal, for the cosmological
perspective see for example
\cite{sunnynew,Oikonomou:2023qfz,Cai:2023dls,Han:2023olf,Guo:2023hyp,Yang:2023aak,Addazi:2023jvg,Li:2023bxy,Niu:2023bsr,Yang:2023qlf,Datta:2023vbs,Du:2023qvj,Salvio:2023ynn,Yi:2023mbm,You:2023rmn,Wang:2023div,Figueroa:2023zhu,Choudhury:2023kam,HosseiniMansoori:2023mqh,Ge:2023rce,Bian:2023dnv,Kawasaki:2023rfx,Yi:2023tdk,An:2023jxf,Zhang:2023nrs,DiBari:2023upq,Jiang:2023qbm,Bhattacharya:2023ysp,Choudhury:2023hfm,Bringmann:2023opz},
see also
\cite{Schwaller:2015tja,Ratzinger:2020koh,Ashoorioon:2022raz,Choudhury:2023vuj,Choudhury:2023jlt,Choudhury:2023rks,Bian:2022qbh}
for earlier works in this perspective, and also
\cite{Guo:2023hyp,Yang:2023aak,Machado:2018nqk} for the axion
description. Although it is not certain whether the 2023 NANOGrav
signal has a cosmological or astrophysical source, or even a
combination of the two, there are theoretical hints toward the
cosmological description. For the astrophysical description of the
signal, see the recent review \cite{Regimbau:2022mdu}. Currently
there are some issues that render the cosmological description
more plausible, although in principle the signal may be a hybrid
of astrophysical and cosmological origin. The obstacles for the
astrophysical description are firstly the lack of a concrete
solution for the last parsec problem \cite{Sampson:2015ada},
secondly, the absence of large anisotropies in the NANOGrav signal
\cite{NANOGrav:2023gor} and thirdly the complete absence of
isolated supermassive black hole merger events. Clearly, the
detection of large anisotropies in the future may point out that
the astrophysical description is plausible
\cite{Sato-Polito:2023spo,Ding:2023xeg}. However, some hint in
this direction should be present even in the 2023 data, so the
near future will shed light on these issues. On the other hand,
there are a lot of cosmological scenarios that may explain the
signal, like phase transitions, cosmic strings, primordial black
holes and so on. With regard to the inflationary perspective, it
seems that ordinary inflation cannot explain in any way the 2023
NANOGrav signal, unless a highly blue-tilted tensor spectral index
is produced and a significantly low-reheating temperature is
required
\cite{sunnynew,Oikonomou:2023qfz,Benetti:2021uea,Vagnozzi:2020gtf}.
In principle, a blue-tilted tensor spectral index can be generated
in the context of many theories, like string gas theories
\cite{Kamali:2020drm,Brandenberger:2015kga,Brandenberger:2006pr},
or some Loop Quantum Cosmology scenarios can also predict such a
tensor spectrum
\cite{Ashtekar:2011ni,Bojowald:2011iq,Mielczarek:2009vi,Bojowald:2008ik},
and moreover the non-local version of Starobinsky inflation
\cite{Calcagni:2020tvw,Koshelev:2020foq,Koshelev:2017tvv} which
also yield an acceptable amount of non-Gaussianities. Furthermore,
it is important to note that conformal field theory can yield a
blue-tilted tensor spectrum \cite{Baumgart:2021ptt}. From the
aforementioned examples, the only which yields a strongly
blue-tilted tensor spectral index is the non-local Starobinsky
model, however in such a case a low reheating temperature is also
needed. In the case that the reheating temperature is actually too
low, the electroweak phase transition is put into peril, since it
cannot be realized thermally, it is required that the reheating
temperature is at least 100$\,$GeV. We shall discuss this
important issue at a later section. Thus the situation is somewhat
perplexed since it is not certain what causes the 2023 NANOGrav
signal. Only the combination of data coming from all the future
experiments like LISA and the Einstein Telescope, including
NANOGrav, may point out toward to a specific model behind the
stochastic signal. This synergy between experiments may also help
determining the reheating temperature.

In view of the above line of research, in this work we shall
report an intriguing result in the context of inflationary
theories and their ability to explain the 2023 NANOGrav signal.
Our analysis requires a stiff, or nearly stiff pre-CMB era, which
can be caused by various mechanisms, see for example
\cite{Co:2021lkc,Gouttenoire:2021jhk,Giovannini:1998bp,Oikonomou:2023bah,Ford:1986sy,Kamionkowski:1990ni,Grin:2007yg,Visinelli:2009kt,Giovannini:1999qj,Giovannini:1999bh,Giovannini:1998bp}.
The approach of Ref. \cite{Co:2021lkc} remarkably fits the line of
research we shall adopt in this work, and the pre-CMB stiff or
kination era is caused by the axion. Actually the authors perform
a thorough and scientifically sound analysis to put constraints on
such a pre-CMB kination era caused by the axion. We shall not use
a specific model in our work, however the constraints of
\cite{Co:2021lkc} are taken into account in our analysis. Also it
is noticeable that recently another work which invokes a stiff era
and the possibility of explaining the NANOGrav 2023 signal
appeared in the literature \cite{Harigaya:2023pmw}, in a different
approach though, compared to the one we shall present in this
work. Coming back to our case, the stiff era is required to
correspond at a wavenumber range $k=0.4-0.9\,$Mpc$^{-1}$, so just
before the recombination era, and recall that the CMB pivot scale
is at $k=0.002$$\,$Mpc$^{-1}$. Note that the CMB modes with
$k=0.002$$\,$Mpc$^{-1}$ reentered the horizon at $z=1100$ so
during recombination, so the stiff era is assumed to have occurred
before the recombination so when $k=[0.4,0.9]\,$Mpc$^{-1}$ hence
for a redshift $z>1100$. The inflationary era can be described by
a standard red-tilted or a blue-tilted theory. With regard to the
red-tilted inflationary era, it can be described by a standard
inflationary scenario or some modified gravity
\cite{reviews1,reviews2,reviews3,reviews4,reviews5,Sebastiani:2016ras,reviews6},
while the blue-tilted inflationary era can be described for
example by an Einstein-Gauss-Bonnet gravity. The striking new
result is that the tensor spectral index in the latter case is not
required to be strongly blue-tilted, but it can be of the order
$\sim \mathcal{O}(0.37)$, which can be easily generated by an
Einstein-Gauss-Bonnet theory. In the case of a blue-tilted
inflationary era, the compatibility with the NANOGrav result comes
easily when the reheating temperature is of the order $T\sim
\mathcal{O}(0.1)\,$GeV, while the red-tilted inflationary case
scenario is not detectable by the NANOGrav, however the prediction
is that can be either detected by LISA or by the Einstein
Telescope, not simultaneously though, depending on the reheating
temperature and the total equation of state (EoS) parameter
post-inflationary.

Regarding the stiff pre-CMB era, we shall assume an agnostic
approach without proposing a model for this, except for the last
section where we shall discuss a potentially interesting model.

 The methodology of reconstruction that will be employed below is
based on the work of \cite{Nojiri:2009kx}. Before we begin any
sort of theoretical calculation it is important to note that we
are working on the well known FLRW metric:
\begin{equation}\label{metric}
    ds^2 = -dt^2+a(t)\sum_{i=1,2,3} dx^2_{i} .
\end{equation}
Now consider the action integral for an $f(R)$ gravity
\begin{equation}
    S = \frac{1}{2k^2}\int  f(R)\sqrt{-g} \,dx^4 ,
\end{equation}
with k being the inverse of the Planck mass, as given by the
relation $k = \frac{1}{M_{pl}}$. Now, varying the action with
respect to the inverse metric tensor we obtain the
equations,\cite{Odintsov:2020th}
\begin{equation} \label{starteq}
    f_R(R)R_{\mu\nu}-\frac{1}{2}f(R)g_{\mu\nu}+(g_{\mu\nu} \Box - \nabla_{\mu}\nabla_{\nu})f_{R}(R) = kT_{\mu\nu} .
\end{equation}
The energy-momentum tensor is given by the equation:
\begin{equation}
    T_{\mu\nu}=\frac{2}{\sqrt{-g}}\frac{\delta(\sqrt{-g}L_m)}{\delta g^{\mu\nu}} ,
\end{equation}
With $L_m$ being the Lagrangian of matter, and the expressions
$f_R$ and $f_{RR}$ represent the first,and second derivative of the function $f(R)$ with respect to the Ricci scalar $R$ ,respectively. The above
relations when evaluated using the FLRW metric [\ref{metric}], and
knowing that the Ricci scalar can be expressed as:
\begin{equation}\label{ricci}
    R = 6\dot{H}+12H^2 ,
\end{equation}
produce the following useful equation:
\begin{equation}\label{Fried}
     -\frac{f(R)}{2}+3(H^2+\dot{H})f_R-18(4H^2\dot{H}+H\ddot{H})f_{RR} +\kappa^2\rho =0 \quad
     \text{(Friedmann equation)} ,
 \end{equation}
which we can recognize as the Friedmann equation in the case of Einstein gravity \cite{Friedmann:1924bb}. To begin the reconstruction technique, we shall make the
following steps to better accomplish constructing a differential
equation in terms of $R$, $f(R)$ and its derivatives. We shall
work in number of e-folds , defined as $N=log(\frac{a}{a_0})$,
with the current scale factor $a_0$ being normalized to unity.
Also, we define the function $G(N) = H^2$. The Ricci scalar
\cite{Nojiri:2009kx}, thus becomes, $R = 3G'(N) + 12G(N)$ ,with $G$ prime
denoting the first derivative with respect to the number of
e-folds, the G function can be written as:
\begin{equation}
    G(N)=\frac{2 n^2 e^{-\frac{2 \text{N}}{n}} \left(b^{1/n}-e^{\text{N}/n}\right)^2}{b^2} ,
\end{equation}
using that fact we can solve the number for the number of e-folds
in terms of the Ricci scalar. As the final expression is really
cumbersome to work with, we shall make the large curvature
approximation. The curvature at primordial times is orders of
magnitude larger than the other parameters and thus, we can make
an expansion of N(R) around infinity. Keeping only the important
terms we get the expression:
\begin{equation}
   N = n \log \left(\frac{2 \left(-\frac{\sqrt{3} b^2 \sqrt{-n} R \sqrt{-2 b^2 n R+b^2 R-3 n}}{\left| 24 n^2-b^2 R\right| }+\frac{24 \sqrt{3} n^2 \sqrt{-n} \sqrt{-2 b^2 n R+b^2 R-3 n}}{\left| 24 n^2-b^2 R\right| }+12 n^2-3 n\right)}{24 n^2-b^2 R}\right)+\log (b) .
\end{equation}
By, invoking a second cosmological equation, i.e the energy-conservation of the cosmological fluids, with a constant EoS parameter the total density can be written as:
\begin{equation}
    \rho_{fluids} = \sum_{i} \rho_{i0}a_{0}^{-3(1+w_{i})}e^{-3(1+w_{i})N}
\end{equation}
Substituting this equation in the Riemann equation we
deduce the following differential equation:
\begin{multline}
9G(N(R))[4G'(N(R))+G''(N(R))]F''(R) + (3G(N) + \\
    \frac{3}{2}G'(N(R)))F'(R) - \frac{1}{2}F(R)+\kappa^2\sum_{i} \rho_{i0}a_{0}          ^{-3(1+w_{i})}e^{-3(1+w_{i})N(R)} = 0 .
\end{multline}
Our analysis, is focused on a field-free reconstruction process, and as such the part containing cosmic fluids will be omitted. Although it is of high interest to point out that this methodology, agrees with the results of \cite{Raychaudhuri:1953yv}-\cite{Choudhury:2019zod}, that approach the problem with the Raychaudhuri equation as the starting point of their analysis. Of course, our results coincide ,on account of the fact that the three equations that govern the dynamics of Cosmology (Friedmann, Raychaudhuri and continuity equation) are not linearly independent,and so by combining any two of them it is possible to derive the third one. In our case, the resulting differential equation, can be simplified if we assume the large curvature
approximation, and keep only the most important polynomial terms
that dominate during primordial times, as the rest of the
expansion is in terms of inverse power of $R$. The differential
equation and coefficients can be seen below:
\begin{equation}\label{diffeq}
    A(R)F''(R)+B(R)F'(R)-\frac{1}{2}F(R)=0 ,
\end{equation}
with:
\begin{equation}
    A(R)=\frac{R^2}{2 n-1} ,  \ B(R) = \frac{(n-1) R}{2 (2 n-1)}.
\end{equation}
Solving the differential equation we obtain the form of the
reconstructed gravity, that reproduces an emergent universe
scenario, namely the f(R) model is:
\begin{equation}
    f(R) = c_1 R^{\frac{1}{4} (- \sqrt{{n^2+10 n+1}}-n+3)}+c_2R^{\frac{1}{4} (+\sqrt{{n^2+10 n+1}}-n+3)} ,
\end{equation}
with $c_1, \ c_2$ being the constants of integration. One notable
aspect of the above expression is that f(R) gravities of the form
$c_1R^{m_+}+c_2*R^{m_-}$ appear often in reconstruction
procedures. Also in order to further support our argument of the
large curvature approximation, we shall present the order of
magnitude of the Ricci scalar during primordial times. The formula
$R = 12H^2+6\dot{H} \approx 12H^2$.The authors of
\cite{Appleby:2009uf} find the scale of inflation to be $H \approx
O(10^{21}) eV$ , therefore the Ricci scalar is $R \approx
O(10^{42}) eV$. It easily follows from the sufficiently large
value of curvature (which is going to be orders of magnitude
larger when the Universe is in its Einstein static phase),
justifies the large curvature approximation as the higher order
terms dwarf the rest of the expression in the series. Also, an inspection of the form of the $f(R)$ gravity reveals as stated in \cite{Oikonomou:2020qah}, the terms of the equations can be grouped as follows:
\begin{equation}
    3*H^2 = \kappa^2 \rho_{tot}, \quad -2\dot{H}=\kappa^2(\rho_{tot} + P_{tot}) ,
\end{equation}
with $\rho_{tot}= \rho_{radiation}+\rho_{gravity}$ and $P_{tot}=
P_{radiation}+P_{gravity}$. The pressure and density due to
gravity are purely geometric terms and are described by the
equations:
\begin{equation}
    \kappa^2 \rho_{gravity} = \frac{F_{R}R-F}{2}+3H^2(1-F_{R})-3H\dot{F_{R}}, \quad \kappa^2 P_{gravity} = \ddot{F_{R}}-H\dot{F_{R}}+2\dot{H}(F_{R}-1)-\kappa^2 \rho_{gravity}
\end{equation}
It is then clear that, as the cosmological fluid is a perfect one, the gravitational pressure and density must follow the same continuity equation, namely:
\begin{equation}
    \dot{\rho}_{gravity}+3H(\rho_{gravity} + P_{gravity}) = 0 ,
\end{equation}
and as such our $f(R)$ model is conserving.

\section{Primordial Curvature Dynamics}\label{sec3}
This section will be dedicated to deriving appropriate expressions for the parameters of that dictate the primordial profile for the scalar and tensor perturbations. The expression mentioned below are the same for inflationary mechanics, but due to the generality of their derivation they can be used to check the viability of our model. The methodology that will be presented below consists of writing easily manipulated expressions for the parameters of the emergent epoch, using them to calculate the spectral indices for scalar and tensor perturbations, as well as the tensor-to-scalar ratio, and lastly compare them to the results demonstrated by the Planck 2018 data ,\cite{Planck:2018vyg}\\
The main assumptions of inflation are that $\ddot{H}<<H\dot{H}$,
and $\frac{\dot{H}}{H^2}<<1$ and the initial emergent epoch is
described by the parameters:
\begin{equation}
\epsilon_1=-\frac{\dot{H}}{H^2}, \quad \epsilon_2 = 0, \quad
\epsilon_3=\frac{\dot{f}_R}{2Hf_R}, \quad \epsilon_4
=\frac{\ddot{f}_R}{H\dot{f}_R} .
\end{equation}
Thus we get from  \cite{Nojiri:2008nt},\cite{Hwang:2005hb} ,the
expressions for the observable quantities:
\begin{equation}
nS = 1+\frac{-4\epsilon_1-2\epsilon_2+2\epsilon_3}{1-\epsilon_1},
\quad r=48\frac{\epsilon_3^2}{(1+\epsilon_3)^2}, \quad nT =
-2(\epsilon_1+\epsilon_3) .
\end{equation}
By far the easier parameter to calculate is $ \varepsilon_1 $,
namely using the scale factor \ref{scalefactor} we get:
\begin{equation}
\epsilon_1= \frac{b^2 e^{\frac{2 \sqrt{2} t}{b}}
\left(e^{-\frac{\sqrt{2} t}{b}}+1\right)^2 \left(\frac{2 n
e^{-\frac{2 \sqrt{2} t}{b}}}{b^2 \left(e^{-\frac{\sqrt{2}
t}{b}}+1\right)^2}-\frac{2 n e^{-\frac{\sqrt{2} t}{b}}}{b^2
\left(e^{-\frac{\sqrt{2} t}{b}}+1\right)}\right)}{2 n^2} .
\end{equation}
Now will try to express the other parameters as a function of
$\varepsilon_1$. From the known literature we can find that,
\begin{equation}
\epsilon_1 = -\epsilon_3(1-\epsilon_4) \leftrightarrow \epsilon_3
= \frac{\epsilon_1}{\epsilon_4-1} .
\end{equation}
The last parameter $\varepsilon_4$ is a little bit trickier to
calculate, namely :
\begin{equation}
\epsilon_4 = \frac{\ddot{f}_R}{H\dot{f}_R} =
\frac{f_{RRR}\dot{R}^2+f_{RR}\ddot{R}}{Hf_{RR}\dot{R}} .
\end{equation}
But, for the time derivative of the Ricci scalar we get after
assuming that $\ddot{H}<<\dot{H}H$ (namely the first inflation
condition):
\begin{equation}
\dot{R}=\frac{d}{dt}(12H^2+6\dot{H})=24\dot{H}H+6\ddot{H} \approx
24\dot{H}H = -24H^3\epsilon_1 .
\end{equation}
Now calculating the time derivative for $\epsilon_1$ ,
\begin{equation}
\dot{\epsilon_1}=\frac{d}{dt}\frac{-\dot{H}}{H^2}=\frac{-\ddot{H}H^2+2H\ddot{H}^2}{H^4}=\frac{-\ddot{H}}{H^2}+\frac{2\dot{H}^2}{H^3}
.
\end{equation}
After ignoring the second order derivatives of $H$ , we
approximately get:
\begin{equation}
\dot{\epsilon_1}=\frac{2\dot{H}H}{H^4}.
\end{equation}
Also setting the function $x(R)=-48\frac{f_{RRR}H^2}{f_{RR}}$, we
get the final expression for the $\epsilon_4$:
\begin{equation}
\epsilon_4=\frac{x(R)\epsilon_1}{2}-3\epsilon_1+2\frac{\epsilon_1^2H}{H\epsilon_1}=\epsilon_1(\frac{x(R)}{2}-1)
.
\end{equation}
\qquad The function $x(R)$, given in terms of the cosmic time,
takes enough space in this letter to render it unreadable. For
this reason , it could be computed either manually by the reader
or seen at the attached Mathematica Notebook, provided in the
documentation \cite{EUS-link}. \par We wish to calculate the
observational parameters, using the number of e-folds. To do that
we integrate the expression $N = \int_{t_i}^{t_{early}} H(t) \,dt
$ , and thus we get the formula for the time of the first horizon
crossing, in terms of the Number of e-folds(N). The time that the
initial H conditions start to break down is when  $\varepsilon_1
=1$ , that happens at time,  $t_{\text{final}} = \frac{b \log
(-n)}{\sqrt{2}}$. That is the reason n had to be negative, so that
the final time is not imaginary. Now the integral is solved with a
straightforward u-substitution, so that the time of exit in terms
of the number of e-folds is:
\begin{equation}
  t_i = \frac{b \log \left(\frac{1}{\frac{(n-1) e^{-\frac{\text{N}}{n}}}{n}-1}\right)}{\sqrt{2}} .
\end{equation}
\qquad The functions $n_S(t)$, $n_T(t)$ and $r(t)$, evaluated at
the time of initial curvature perturbations(ti), contain 5 free
parameters. The order of the parameters is [N,n,b,$c_1$,$c_2$].
Its obvious by now that the functions in question are extremely
large, hundreds or thousands of characters in length and as such
cannot be displayed in this letter. The process of finding the
correct range of parameters was made extremely easy by utilizing
the Genetic Algorithm showcased at [\textbf{Appendix}]. In general
for $b>1000$ the parameters don't present significant change,
while being highly influenced by the value of $n$. Therefore for
50-60 e-folds, and $b$ larger than 1000,the n value that makes the
model viable is in the range of -30 to -60. An example is that the
point [60, -60 ,1000, 1,1 ] produces $nS$ = 0.9630737 and $r$ =
0.0042824. These values comfort with the Planck 2018 constraints:
\begin{equation*}
    nS = 0.962514 \pm 0.00406408 , \quad r <0.064
\end{equation*}
\qquad It must be noted that the parameters $c_1$ and $c_2$ play
no role in the viability of the model, as with the substitution
mentioned earlier the ratio $\frac{f_{RRR}}{f_{RR}}$ found in
$\varepsilon_4$ is not strongly dependant on the constants of
integration $c_1, \ c_2$. Also the range of viability of the model
naturally confirms the demand that b must be positive and, so that
the scale does not dive into the negatives, while also having n
smaller than zero for the universe to experience rapid
development. The nT results remain of high interest as for some
inputs of the free parameters the nT becomes negative and as such
the results of our model will not be detected by the next
generation of gravitational waves detectors.\par
\begin{figure}[h]
    \centering
    \includegraphics[width=8cm]{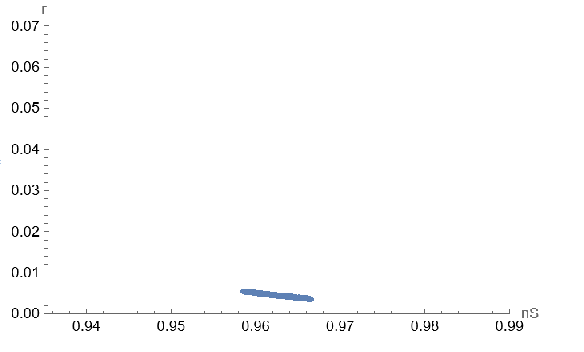}
    \caption{The r-nS diagram for the viable points.It fits quite comfortably inside the range specified by Planck}
\end{figure}
The comoving Hubble Horizon represents the maximum distance
information, travelling at the speed of light, can reach us. The
mathematical expression is given by:
\begin{equation}
    R_h(t)=\frac{1}{a(t)*H(t)} .
\end{equation}
In our model, this takes the form:
\begin{equation}
    R_h(t) = -\frac{e^{\frac{\sqrt{2} t}{b}} \left(e^{-\frac{\sqrt{2} t}{b}}+1\right)^{1-n}}{\sqrt{2} n} .
\end{equation}
As presented in the plot below, the comoving Hubble radius
plummets during the accelerated phase of our emergent universe
scenario, a key characteristic that justifies treating that phase
using inflationary dynamics.
\begin{figure}[h]
    \centering
    \includegraphics[width=9cm]{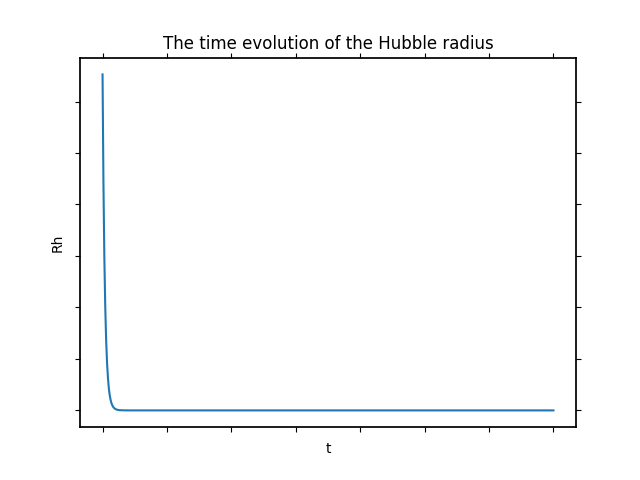}
    \caption{The Hubble radius in terms of time, for n=-60 and b=2000}
\end{figure}
\quad Now that we have been introduced to the indices
$\varepsilon_i , i=1,3,4$, there must be a discussion regarding
the incompatibility of mainstream scale factors to the standard
procedure of inflationary dynamics. Namely for scale factors of
the form $a(t) = b(1+e^{t})^n$, fulfill the exit condition
$\varepsilon_1 =1$, when n is negative, but such a value for n
contradicts the main idea of emergent universes for the scale
factor plummets to zero instead of showing a rapid growth. Same
problems appear when introducing more free parameters into the
scale factor as the exit condition demands that the free variables
make an unnatural scale factor inconsistent with the demands of an
emergent universe.The same issues are presented in different scale
factors such as: $a(t) = b*\text{ln}(k+e^{t*p})$. These issues
where the driving force behind the choice of the peculiar scale
factor that we introduced at the start of the letter.\par

\section{Big Bang Nucleosynthesis constraints }\label{sec4}
The following methodology relies on the work of
\cite{Asimakis:2021yct},\cite{Anagnostopoulos:2022gej}. Having
obtained the input parameters that fall in line with the Planck
2018 data concerning nS and r, we shall examine the limitations
being put in place by BBN data. First of all, the model we
presented was not suitable for such an investigation, by adding a
term of +R , we reach the new form of:
\begin{equation}
    f(R) = R + _1 R^{\frac{1}{4} (- \sqrt{{n^2+10 n+1}}-n+3)}+c_2R^{\frac{1}{4} (+\sqrt{{n^2+10 n+1}}-n+3)} ,
\end{equation}
at first such a change may seem arbitrary, but there are valid
grounds for such an addition. Specifically due to the form our
differential equation \ref{diffeq}, the residue of substituting
the extra R term back in, is of order 0.1*R. That deviation is
absolutely acceptable if we consider the general large curvature
approximation that was used to deduce the form of the f(R)
gravity. On the other hand, that term may not reproduce the scale
factor introduced at the start of our letter, but a modified
version of it, so that it still has the properties discussed.
Another important thing to note is that our model for the viable n
range takes the form $f(R)=c_1*R^m+c_2*R^k$, with $m \approx 2$
and $k \approx 20$, meaning it represents an enhanced Starobinksy
model. As we will see later the $c_2$ constant is completely free
meaning no matter what value it takes the viability of model is
unchanged, but due to obvious physical arguments, at the end of
our analysis we shall set equal to zero. Regarding the
introduction of the R term it must be noted ,that because the
$\varepsilon_3$ index depends only on the second and third
derivative of the f(R) gravity in terms of the Ricci scalar, it
does not spoil our process of finding a viable range of parameters
using the nS and r parameters. \par After exiting the inflationary
phase, our Universe enters a radiation dominated era in which
after the first few seconds the  interaction between quarks
freezes out and begins the process of making the nuclei of heavier
elements and their isotopes,from protons and neutrons, in a
process called Nucleosynthesis. Considering that due to adiabatic
expansion the equality describing entropy $S=T^3*a^3$ must not be
violated, the temperature drops as $T \propto 1/a$. That small
window of time in the order of a few minutes, and concerning
energies of MeV to GeV, is really well understood as such energy
scales have been studied and replicated in large particle
colliders, resulting in some of the most precise observations on
the field of Cosmology.\par The Standard model of Cosmology
($\Lambda CDM$) , guided by the Friedmann equation in the scope of
General Relativity, describes the expansion of the universe in
terms of time, during BBN, as:
\begin{equation}
    H^2_{GR} = \frac{M^{-2}_{P}}{3}\rho_r ,
\end{equation}
with $\rho_r$ being the density of radiation that dominates during
BBN. For relativistic species it is given in the form of :
\begin{equation}
    \rho_r = \frac{\pi^2}{30}g_{\star}T^4 ,
\end{equation}
with $g_{\star} \approx 10$ being the degrees of freedom of the relativistic species. Also it should emphasized that the constant $M_P$ deviates from the Planck Mass, as $M_{Pl}=\sqrt{8\pi}M_{P} = 1.22 * 10^{19}$ GeV \\
During BBN the total rate of conversion between protons to
neutrons, due to interactions with the weak force, can be
described by adding the rates of each of the subsequent reactions
taking place. To be precise, the neutron conversion to proton
reaction rate is:
\begin{equation}
    \lambda_{pn}(T) = \lambda_{n + e^{+} \rightarrow p + \bar{\nu_{e}}
}+ \lambda_{n + \nu_{e} \rightarrow p + e^{-}} +\lambda_{n
\rightarrow p + e^{-}+\bar{\nu_{e}}} ,
\end{equation}
along with the inverse reactions we get the total rate
$\lambda_{total}(T)=\lambda_{pn}+\lambda_{np}$. Assuming that the
energies are high enough for Boltzmann distribution to be able to
describe the present processes, while the species are highly
relativistic so that their masses are insignificant in comparison
to their energies, we can derive the simplified expression:
\begin{equation}
\lambda_{total}(T) = 4AT^3(4!T^2+2*3!QT+2!Q) ,
\end{equation}
with Q being the difference between the masses of the two
nucleons, and equal to $1.29*10^{-3} GeV$, and A a constant equal
to $1.02 * 10^{-11} GeV^{-4}$. The Fermi-Dirac distribution given
by: $p(T,\varepsilon_i)=
\frac{1}{1+\exp{(\varepsilon_i-Q)/(k_B*T)}}$. As
$\varepsilon-Q>>kT$ during this epoch, we can approximate it as
$1+\exp{(\varepsilon_i-Q)/(k_B*T)} \approx
\exp{(\varepsilon_i-Q)/(k_B*T)}$ leading back to Maxwell-Boltzmann
statistics. As the Universe expands, and the reaction rate of this
equation gets smaller and smaller as temperature drops, at some
specific temperature $T_{freeze}$ , the reaction freezes out. That
happens when :$\lambda_{total}(T) \approx 1/H$ leading to a freeze
out temperature of:
\begin{equation}
    T_f = (\frac{4\pi^3 g_{\star}}{45M^2_{Pl}c^2_q})^{\frac{1}{6}}  \approx 0.6 MeV ,
\end{equation}
the constant $c_q$, being equal to $9.8*10^{-10} GeV^{-4}$. Adding
curvature terms to the standard GR model is going produce a
different rate of expansion meaning the reactions will occur at
different rates, leading to different mass ratios for the nuclei
produced. Being more specific lets consider that our Hubble rate,
after using equation \ref{Fried} for modified gravities yields:
\begin{equation}
    H^2 = \frac{M^{-2}_P}{3}(\rho_r+\rho_{grav}) ,
\end{equation}

with the extra terms acting as the density of a fluid:
\begin{align}
    \rho_{grav} \left(\frac{M_P^{-2}}{3}\right)^{-1} &= \frac{1}{2} \left(-6^{\frac{1}{4} \left(-\sqrt{n^2+10n+1}-n+3\right)}\right) \notag \\
    &\quad \times \left(H'(t) + 2 H(t)^2\right)^{\frac{1}{4} \left(-\sqrt{n^2+10n+1}-n+3\right)} \notag \\
    &\quad \times \left(\text{c1} + \text{c2} 6^{\frac{1}{2} \sqrt{n^2+10n+1}} \left(H'(t) + 2 H(t)^2\right)^{\frac{1}{2} \sqrt{n^2+10n+1}}\right) \notag \\
    &\quad + 3 \left(H'(t) + H(t)^2\right) \notag \\
    &\quad \times \left(\frac{1}{4} \text{c2} \left(\sqrt{n^2+10n+1}-n+3\right) \left(6 H'(t) + 12 H(t)^2\right)^{\frac{1}{4} \left(\sqrt{n^2+10n+1}-n+3\right)-1} \right. \notag \\
    &\quad \left. -\frac{\text{c1} \left(\sqrt{n^2+10n+1}+n-3\right) \left(6 H'(t) + 12 H(t)^2\right)^{-\frac{1}{4} \sqrt{n^2+10n+1}-\frac{n}{4}}}{4 \sqrt[4]{6} \sqrt[4]{H'(t)+2 H(t)^2}}\right) .
\end{align}

Taking the square root of both sides and taking out the radiation
density term we get the expression below, which is going to be
just the GR Hubble rate along with a small deviation $\delta H$.
As can be seen from the bibliography, this is a common approach to
tackle the gravitational effects during Nucleosynthesis. To make
this point more concrete, lets consider the argument we used to
justify the large curvature approximation. From the approximation
$R \approx H^2$, while the universe undergoes a radiation
dominated era after reheating, which dynamics will be left
ambiguous besides the condition that it has short time-span, the
Hubble rate scales as $H \sim t^{-1}$. Taking into consideration
that after reheating the Hubble rate will be orders of magnitude
smaller than the value of $O(21)$ $eV$ , during inflation, and the
many orders of magnitude that describe the timescale between the
end of reheating and the start of BBN, we can safely conclude that
$H << 1$. Finally, the higher curvature terms will be dominated by
the simple Einstein-Hilbert part of gravity model, and will only
act as perturbations. So, we can write the equation:
\begin{equation}
    H=H_{GR}\sqrt{1+\frac{\rho_{grav}}{\rho_r}}=H_{GR}+\delta H.
\end{equation}
Using the above relationship, and taking into consideration that $\rho_r>>\rho_{grav}$, as the gravitational effects are the first order perturbations:
\begin{equation}
    \delta H = H_{GR}(\sqrt{1+\frac{\rho_{grav}}{\rho_r}} -1 ) \\
    \delta H \approx H_{GR}(1+\frac{\rho_{grav}}{2\rho_r}/ -1 )=H_{GR}\frac{\rho_{grav}}{2\rho_r}
\end{equation}
As during BBN, the Hubble rate and the freeze-out temperature are, in leading order, related by:
\begin{equation}
    H = \lambda_{tot} \sim qT^5
\end{equation}
Therefore, a deviation from the $H_{GR}$ will produce an according deviation for the freeze-out temperature, $T_{f}$. These deviations, after combining the previous equations give the final expression:
\begin{equation}\label{bbneq}
    \frac{\delta T_f}{T_f}=\frac{\rho_{grav}}{\rho_r}\frac{H_{GR}}{10c_qT_f^5} .
\end{equation}
The above equation is well determined as it is crucial for the
calculation of the mass fraction of $^4$He. The
observational results determine that it must be:
\begin{equation}
    \frac{\delta T_f}{T_f} <4.7 *10^{-4} .
\end{equation}
Applying the presented methodology to our model, and demanding
that the inequality must be respected we get a really satisfactory
result. The $c_2$ parameter is irrelevant to our analysis, and it
can take any value without spoiling the BBN results while the $c_1$
parameter is restricted by a constant $P$ of $O(128-129)$
depending on the choice of the value of n. That means that for any
value of $-P<c_1<P$, our model does not produce any changes to the
results of BBN, thus omitting the need for fine tuning the
integration parameters and presenting an excellent opportunity for
a viable f(R) model even during late times.

\section*{Concluding Remarks and Discussion}

The present letter, after taking into consideration the
inconsistencies at producing  imaginary exit times and unnatural
models for an emergent universe scenario, explored the viability
of scale factor \ref{scalefactor}. We used the techniques
developed at \cite{Nojiri:2009kx}, and reconstructed the form of
f(R) in the matter-free, flat universe approach that reproduces
that specific scale factor at large curvatures. Using the
inflationary parameters $\varepsilon_i , i=1,2,3$, it was possible
to extract the spectral indices for scalar and tensor
perturbations along with the tensor-to-scalar ratio , and find the
appropriate range for the parameters that satisfies the
constraints set by Planck 2018 data, for 50-60 e-folds.The
parameter b should be at least of order O(3), with the appropriate
choice of n being between -30 to -60, while also noting that the
free constants of integration played no role in the whole
analysis. Going even further, we introduced an extra linear term,
+R that is just the normal expression for General Relativity, and
explored the ability of the model to reproduce BBN constraints.
After carefully going throughout the process presented at
\ref{sec4},we found the constant $c_2$ has no significance on the
results of Nucleosynthesis while the constant $c_1$ can take
virtually any value , negative or positive, up to order of O(128).
The model has the form of an enhanced Starobinksy model, while the
problematic term of large curvature can easily be extinguished
without any issues to our model. Therefore the present f(R) for is
able to reproduce the inflationary period, the BBN along with an
emergent universe scenario that bypasses the quantum gravity and
singularity regime.

\section*{Acknowledgments}

This research has been funded by the Committee of Science of
the Ministry of Education and Science of the Republic of
Kazakhstan (Grant No. AP19674478) (V.K. Oikonomou). G. Kafanelis
is indebted to E.N. Saridakis for his guidance on implementing the
BBN constraints, and also to his fellow colleague Mr Fotis
Fronimos for his ingenious insights and productive conversations.

\section*{Appendix}\label{sec:appendix}

In this section we shall present the genetic algorithm that
enables us to find the viable points efficiently and reliably. The class GenAlgo represents the most vital part of the code base. During initiation the constructor expects the $n_S$ and $r$ functions, that were calculated in the framework of the working theory, the symbols used in them in the form of a \textbf{Sym.symbols} object, and lastly a string that represents the fitness function that will be used during the genetic algorithm. We leverage the speed due to vectorizition that is provided by the \textbf{Sympy.lambdify} object to speed up the numerical calculations. As our group mostly works with Mathematica Notebooks, the object allows raw Mathematica format to be transformed into appropriate \textbf{Sympy} expressions.
\begin{verbatim}
class GenAlgo:
    def __init__(self,str_nS,str_r,input_symbols,mathematica_form=True,
    fitness_expression='devr * devr + devnS * 100 + devN * 0.1'):
        if mathematica_form:
            self.python_expr_r = parse_mathematica(str_r)
            self.python_expr_nS = parse_mathematica(str_nS)
        else:
            self.python_expr_r = str_r
            self.python_expr_nS = str_nS
        self.function_symbols = input_symbols
        self.fitness_expression = fitness_expression

        #The Planck 2018 constaints
        self.rlim = 0.064
        self.nSmin = 0.962514 - 0.00406408
        self.nSmax = 0.962514 + 0.00406408

        #The range of the e-folds
        self.Nmax = 60
        self.Nmin = 50

        self.nSsympyFunction = sym.lambdify(self.function_symbols,self.python_expr_nS,'numpy')
        self.rsympyFunction = sym.lambdify(self.function_symbols,self.python_expr_r,'numpy')

        self.fitness_expression = fitness_expression
        self.scoringSymbols = sym.symbols('devnS devr devN')
        self.scoringFunction = sym.lambdify(self.scoringSymbols,self.fitness_expression,'numpy')
        pass
\end{verbatim}
Each symbol in these expressions represents a parameter of our
model, so for the $n_S$ and $r$ functions to be able to take the
appropriate inputs and produce a numerical result, they are
defined in the form presented below. We used the symbols
\textbf{Ne1 n b c1 c2} to represent the free parameters in the
model of \ref{scalefactor}, always starting with the number of
efolds, but they can be changed to fit other phenomenological
models with different parameters. It must be noted that if the
evaluated expression is imaginary, \textbf{Sympy} will throw an \textbf{Exception} and thus the functions will return a value of
positive infinity, the reason behind this choice will be explained
thoroughly when we introduce the fitness function.
\begin{verbatim}
def nS(self,args):
        try:
            with warnings.catch_warnings():
                warnings.filterwarnings("error")
                var = self.nSsympyFunction(*args)
                return var
        except Exception as e:
            return float('inf')

def r(self,args):
            try:
                with warnings.catch_warnings():
                    warnings.filterwarnings("error")
                    var = self.rsympyFunction(*args)
                    return var
            except Exception as e:
                return float('inf')
\end{verbatim}
We shall break off our explanation of the \textbf{GenAlgo} to introduce another integral component of our code-base. The \textbf{Solution} object holds a \textbf{NumPy} array,
which contains the parameter values in the same order as they were
introduced in the $n_S$ and $r$ functions. Additionally, the
\textbf{Solution} object incorporates a crucial function called
\textbf{score}, serving as the fitness function within our
algorithm. The \textbf{scoringFunction} is passed from the \textbf{GenAlgo} during the computation process. The \textbf{dev\_keyword} calculates the absolute difference of the 'keyword' from its appropriate bounds. The bounds for each 'keyword' is transfered from the \textbf{GenAlgo} object. Along with satisfying the Planck 2018 contraints we impose the condition that the number of e-folds must be in the 50-60 range to appropriate adress the Homogeneity problem. In the
scoring function, we minimally weight the deviation from the constraint in
the $r$ function to diminish its impact on the overall score.
During training, when all deviations were considered equally, the
algorithm tended to excessively prioritize reducing the deviation
from the $r$ constraint since there is no lower bound, thus
resulting in a skewed optimization. Furthermore, we carefully
select the weights so that the deviations in both $n_S$ and $N$
have comparable magnitudes. It's important to note that, contrary
to the typical evaluation of fitness algorithms, our algorithm is
rewarded for achieving lower scores, indicating a higher
likelihood of generating a viable phenomenology within the current
values stored in the \textbf{Solution} object. In order to address
this confusion, we named the function "score" instead of the
standard "fitness". One last comment concerning the \textbf{score}
function is that, as was mentioned earlier the $n_S$ and $r$
functions return positive infinity when the evaluated expression
has an imaginary part, consequently \textbf{Solutions} that
produce non real results are quickly ruled out by the algorithm.
\begin{verbatim}

class Solution:
    def __init__(self, genome):
        self.genome = genome

    def score(self,other):
        nS_value = other.nS(self.genome)
        r_value = other.r(self.genome)
        efolds = self.genome[0]
        dev_nS = abs(nS_value-other.nSmin)+abs(nS_value-other.nSmax)
        dev_r = abs(r_value-other.rlim)
        dev_N = abs(efolds-other.Nmin)+abs(efolds-other.Nmax)
        objectScore = other.scoringFunction(dev_nS,dev_r,dev_N)
        return objectScore
\end{verbatim}
To initiate the algorithm, the function \textbf{genetic\_algo\_start} is invoked. This function accepts several input parameters, including the range for each symbol, the desired population size of \textbf{Solutions} for training, the quantity of the top-performing individuals in each generation that will participate in the crossover for the subsequent generation, and the total number of iterations for which the algorithm will execute. Additionally, if a pre-existing trained population is available, it can be employed as the initial population.. Each population consists of an array
containing \textbf{Solutions}. \par
\begin{verbatim}
    def start(self,bounds, size_population, fit_lim, iterations, loaded_population=None):
        #Making the first population manually
        fit_lim = fit_lim+1
        best_scores=np.zeros((iterations,fit_lim+1),dtype=float)
        pops = np.zeros(iterations+1, dtype='object')
        best_pops = np.zeros(iterations+1, dtype='object')
        population = np.empty(size_population, dtype='object')
        if loaded_population is None:
            for i in range(size_population):
                var_genome_0 = self.generator(bounds)
                sol = Solution(genome=var_genome_0)
                population[i] = sol
        elif isinstance(loaded_population, np.ndarray) and loaded_population.dtype == object:
            population = loaded_population[-1]
        if len(bounds) != len(self.function_symbols):
            raise ValueError(f"Different dimensions.
             Bounds dimensions : {len(bounds)} and total symbols : {self.function_symbols}")
\end{verbatim}
The initial population is constructed using the \textbf{generator} method of the \textbf{GenAlgo} class. This function takes the relevant
bounds, which determine the range for each variable, and generates
an array of the same length. Each element in the array corresponds
to a random value within the specified bounds for the
corresponding variable. To address a potential issue with the
\textbf{random.uniform} command, which tends to under represent
smaller numbers when the bounds differ significantly (up to three
orders of magnitude), we employ a scaling technique. Firstly, we
apply a logarithmic scaling using $log_{10}$ to reduce the bounds.
Then, we generate the random input within this scaled range.
Finally, we scale the generated value back up while preserving the
appropriate sign, and append it to the final array. This approach
ensures a more balanced representation of all possible values for
a specific variable, even when there are substantial discrepancies
between the upper and lower bounds. However, it is worth noting
that this implementation requires non-zero bounds. Nevertheless,
this limitation can be easily resolved by choosing a suitably
small bound value.
\begin{verbatim}
def generator(self,b):
        n = len(b)
        output = np.zeros(n)
        for index, value in enumerate(b):
            lower_bound,upper_bound = value
            l_b, u_b = (math.log10(abs(lower_bound)) , math.log10(abs(upper_bound)) )
            cons_sign = -1 if lower_bound < 0 or upper_bound < 0 else 1
            output[index] = (10**(random.uniform(l_b,u_b)))*cons_sign
        return output
\end{verbatim}

Next, begins the process of calculating the score for
each player in the current generation. The scores of the \textbf{Solutions} are stored in the
\textbf{scoreboard} array. By sorting the \textbf{scoreboard}
array we find the indices of the \textbf{fit\_lim} lowered
scored players, and create a new array that contains those
solutions, called \textbf{best\_candidates}, and append it to the
larger array \textbf{best\_pops}
\begin{verbatim}
    for generation in range(iterations):
        #Initializing
        t1 = time.time()
        scoreboard = np.zeros(size_population)
        population = pops[generation]

        #Scoring
        scoreboard = parallel(population)

        best_candidates_pos = np.argpartition(scoreboard, fit_lim+1)[:fit_lim+1]
        #Fitness
        best_candidates = np.zeros(fit_lim+1,dtype='object')
        index_position = 0
        for position in best_candidates_pos:
            var = population[position]
            best_candidates[index_position] = var
            index_position = index_position +1
        best_pops[generation]=best_candidates.copy()
\end{verbatim}
During the subsequent phase of the algorithm, we evenly
distribute the \textbf{genome} (the array of values) of each of
the top-performing players into a temporary array called
\textbf{genetic\_tree}. To illustrate this, let's consider an
example where our population consists of 1000 elements, and we
choose to retain the 5 best players. In this scenario, the first
200 elements of the \textbf{genetic\_tree} array would correspond
to the genome of player 0, the next 200 elements would correspond
to the genome of player 1, and so on. By doing this, we ensure
that all players have an equal representation in the
\textbf{genetic\_tree}, placing them on a level playing field for
the next steps of the algorithm.\par
\begin{verbatim}
    #Crossover
    for indicator in range(len(ranges)-1):
        lower = math.floor(ranges[indicator])
        upper = math.floor(ranges[indicator+1] )
        var_genome = best_candidates[indicator].genome
        for index in range(int(lower),int(upper),1):
            genetic_tree[index]=var_genome
\end{verbatim}
 On to the stage of mutation. The \textbf{mutate} function,
takes as input the \textbf{gene} of a \textbf{genome}, adds a
random value between (-\textbf{gene}/s,\textbf{gene}/s) and
returns the modified \textbf{gene}. The value of $s$ is set to 1,
as we found through experimentation that it leads to a quicker and
more stable convergence rate.
\begin{verbatim}
    def mutate(self,gene): # The mutating function
        self.scale = 1
        return gene + np.random.uniform(-gene/self.scale,gene/self.scale)
\end{verbatim}
After creating a \textbf{new\_population} that contains the
mutated \textbf{genomes} of the \textbf{best\_candidates}, we
replace the first elements of the array the \textbf{Solutions}
with \textbf{best\_candidates}. The reason behind this execution
is that due to pure randomness the next generation can score
higher than the current generation, as the \textbf{genomes} can
mutate away from the local optimum. By incorporating the
\textbf{best\_candidates} into \textbf{new\_population}, we
guarantee the best players of the next generation score as low as
the best players of the current one. Lastly we append
\textbf{new\_generation} into \textbf{pops}.
\begin{verbatim}
    #Mutate
    new_population = np.array([Solution(np.array([mutate(gene) for gene in tree])) for tree in genetic_tree])
    #Keeping the best players of the previous generation
    for i in range(len(best_candidates)):
        new_population[i]=best_candidates[i]

    pops[generation+1] = new\_population
\end{verbatim}
 The function returns the arrays \textbf{best\_pops} and
\textbf{pops}. In the notebook
found at, \cite{EUS-link}, we use the \textbf{clear} and
\textbf{get\_values} functions, that in conjunction return the
truly viable ranges found by the algorithm along with the values
of the $n_S$ and $r$ functions that they produce.
\begin{verbatim}
    def process_point(self,obj):
        genome = obj.genome
        test_nS = self.nS(genome)
        test_r = self.r(genome)
        if self.nSmin < test_nS < self.nSmax and test_r < self.rlim:
            return [test_nS, test_r]
        else:
            return [None,None]

    def clear_points(self,arr):
        flattened = np.concatenate(arr).ravel()
        values = []
        nS_arr = []
        r_arr = []
        for obj in flattened:
            returnValue = self.process_point(obj)
            if returnValue[0] is not None:
                values.append(obj.genome)
                nS_arr.append(returnValue[0])
                r_arr.append(returnValue[1])
        return values, nS_arr, r_arr
\end{verbatim}
 To speed up the numerical calculations and bypass the GIL, the \textbf{RunParallel} function can be invoked that runs many populations in parallel using the \textbf{Pathos} module for a single \textbf{GenAlgo} object. It maps the provided list of arguments to the \textbf{ParallelStart} method in the \textbf{GenAlgo} object. Below the specifications for each function are presented. The \textbf{RunParallel} function utilizes a different way of clearing the points, and as such returns the total score accounting for all the parallel populations, along with all the calculated viable points, and their $n_S$, $r$ values.
\begin{verbatim}
def RunParallel(obj,argsList):
    numCPUS = pathos.helpers.cpu_count()
    pool = ProcessingPool(nodes=numCPUS)

    args_list = argsList

    resultsParallel = pool.map(obj.ParallelStart, args_list)

def ParallelStart(self,args):
        bounds, size_population, fit_lim, iterations = args
        loaded_population = None
\end{verbatim}

 By running the algorithm with the appropriate forms of our
$n_S$ and $r$ functions, that where extracted from the Mathematica
Notebook, with the \textbf{RunParallel} function, and the following parameters
\begin{verbatim}
input_symbols = sym.symbols('Ne1 b n c1 c2')
fit_expr = 'devr * 0.1 + devnS * 50 + devN * 0.1
set_bounds = np.array([(50,60),(-10**(20),-10**(-20)),
(-10**(20),+10**(-20)),(-10**(20),10**(20)),(-10**(20),10**(20))])
args_list = [(set_bounds,5000,20,15)]*16
\end{verbatim}
 The algorithm after running for about 1 minute , managed to calculate a total of 1.2 million points and produce the following diagram for the values of $n_S$ and $r$, with 5086 viable points.

\begin{figure}[H]
    \centering
    \includegraphics[scale = 0.75]{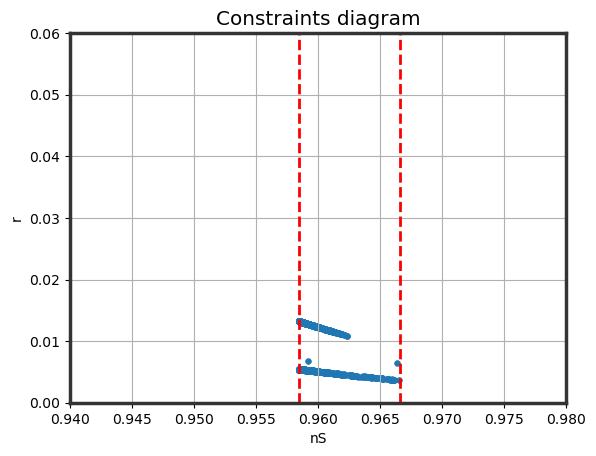}
    \caption{$n_S$-$r$ diagram for the viable points found by the algorithm}
\end{figure}
As it is evident, the algorithm is extremely fast and efficient at finding the optimum set of parameters to fit our constraints, all while being extremely user and developer friendly. We hope the algorithm is used in the work of other inflation researchers as it automates a rather tedious and time consuming
task. Any feedback or suggestions for improving the algorithm are
more than welcome, and we can be contacted through the emails
specified at the start of the letter. \\

\end{document}